# Measuring the Fidelity of Asteroid Regolith and Cobble Simulants

Philip T. Metzger, Florida Space Institute, University of Central Florida, Orlando, Florida, USA, Philip.T.Metzger@ucf.edu

Daniel T. Britt, Department of Physics, University of Central Florida, Orlando, Florida, USA

Stephen Covey, Deep Space Industries, Inc., 6557 Hazeltine National Dr., Orlando, FL 32822

Cody Schultz, Department of Physics, University of Central Florida, Orlando, Florida, USA

Kevin M. Cannon, Department of Physics, University of Central Florida, Orlando, Florida, USA

Kevin D. Grossman, Granular Mechanics and Regolith Operations, Engineering and Technology Directorate, NASA Kennedy Space Center

James G. Mantovani, Granular Mechanics and Regolith Operations, Engineering and Technology Directorate, NASA Kennedy Space Center

Robert P. Mueller, Granular Mechanics and Regolith Operations, Engineering and Technology Directorate, NASA Kennedy Space Center

## Abstract

NASA has developed a "Figure of Merit" method to grade the fidelity of lunar simulants for scientific and engineering purposes. Here we extend the method to grade asteroid simulants, both regolith and cobble variety, and we apply the method to the newly developed asteroid regolith and cobble simulant UCF/DSI-CI-2. The reference material that is used to evaluate this simulant for most asteroid properties is the Orgueil meteorite. Those properties are the mineralogical and elemental composition, grain density, bulk density of cobbles, magnetic susceptibility, mechanical strength of cobbles, and volatile release patterns. To evaluate the regolith simulant's particle sizing we use a reference model that was based upon the sample returned from Itokawa by Hayabusa, the boulder count on Hayabusa, and four cases of disrupted asteroids that indicate particle sizing of the subsurface material. Compared to these references, the simulant has high figures of merit, indicating it is a good choice for a wide range of scientific and engineering applications. We recommend this methodology to the wider asteroid community and in the near future will apply it to additional asteroid simulants currently under development.

## I. Introduction

*Extraterrestrial Simulants* are simulated space materials: geological materials of extraterrestrial bodies including crystalline solids in the form of dust, regolith, boulders, and ice. Simulants are used because there is a large gap between the amount of space materials needed for research and technology development and the amount actually available in the meteorite collection or sample return. Simulants are beneficial for more than one reason. The obvious one is that a simulant can replicate particular physical and chemical properties of a space material with sufficient fidelity that it can be used in lieu of the actual space material in technology tests (or other research) to truthfully indicate how the space material would perform in the same situation. Additional benefits are the improved economies of scale and synergies they create between users. Rather than each project developing its own simulant at its own expense, a few community members may develop one and make it a known entity in the community, typically with simple nomenclature like MLS-1 (Weiblen and Gordon, 1988; Weiblen et al., 1990) or JSC-1 (Willman et al., 1995). The developers then mass produce it, characterize it, document it, and distribute it. This division of labor between simulant developers and simulant users enables greater investment in its fidelity, better characterization, and more complete documentation of its properties and performance than most individual projects could afford. Users often perform and publish characterization tests beyond what the developer provided, extending its usefulness so more users adopt it. Likewise, each reported engineering or scientific test with a simulant encourages others to adopt it so their project can compare with the prior results, and this broadening base of users creates a virtuous cycle.

Unfortunately, these benefits have been offset by misunderstanding of the focused nature of simulants. The exotic processes that created some space materials gave them exotic properties that are too expensive to completely replicate, so simulant developers must choose a subset of the properties. For example, the lunar soil simulant JSC-1, which was re-created as JSC-1A (Carpenter et al., 2006), was intended mainly to replicate the particle size distribution and the mechanical behaviors of lunar soil, and its chemistry has general similarity to some Apollo 14 samples, but it does not simulate the mineralogy of typical lunar soil or even mare soil (Taylor and Liu, 2010), nor the spectrum of particle types such as agglutinates (Rickman, Edmunson and McLemore, 2012), the patina with nanophase iron (Hill et al., 2007) and resulting spectral properties (Pieters et al., 2007) and superparamagnetic susceptibility (Liu, et al., 2007; Gaier, 2008), nor most other of the 32 properties of lunar soil identified by a NASA workshop (McKay and Blacic, 1991). As a result, JSC-1A is well suited for some particular categories of testing but not, e.g., for testing the chemical extraction of resources. Many technologists have not understood this important point, and as a result JSC-1A has been used inappropriately (Taylor and Liu, 2010; LEAG, 2010; Taylor et al., 2016). In 2010, a NASA advisory committee reported that this produced a waste of time and money and "can lead to potentially misleading results that could have disastrous consequences resulting in hardware that does not function properly in the actual lunar environment." (LEAG, 2010)

To deal with this, NASA's In Situ Resources Utilization (ISRU) project created a lunar soil simulants team that reviewed the history and requirements of simulants and created a rigorous process to create future simulants (Edmunson et al., 2010; McLemore, 2014). Among other things, this process established a standard method to determine the applicability of simulants to different types of tests. This method is summarized in Figure 1. It utilizes a *Figure of Merit* (FoM) system and a *Fit-To-Use Table* as follow:

1. The team decided what types of applications were of interest (Sibille, et al., 2006); e.g., using the simulant for lunar soil drilling tests, or chemically reducing the minerals in the soil to extract oxygen.

2. The team decided what properties of the soil should be measured, considering how these properties govern the soil's behavior in the important applications (Rickman et al., 2010; Rickman and Schrader, 2010); e.g., bulk friction and cohesion of the regolith determine drill penetration in drilling tests, whereas the mineralogy is important during chemical reduction tests. Note that some properties will be derivative from other properties; e.g., the correct internal friction coefficient should emerge naturally if the particle shapes, sizes, and surface properties are correctly replicated. Thus, the NASA team chose to measure only primary properties instead of derivative ones for the FoM.

3. The team or other members of the community measured those properties of the simulants (Schrader et al., 2009).

4. The team chose particular extraterrestrial samples to use as the reference materials for the FoMs; e.g., they chose the Apollo 16 core 64001/64002 to represent lunar highlands soils.

5. They measured the same properties of the reference material that they had measured for the simulant.

6. They used a formula to compare the simulant's measurements with the reference material's measurements (Schrader and Rickman, 2009). The formula is similar to an "inner product" between unit vectors, so when the measurements perfectly agree the formula evaluates to unity, but when they completely disagree the value is zero, and the usual case is somewhere between. This value is the FoM. There will be one FoM for each measured property: one for particle sizing, another for mineralogy, etc. One set of all these FoMs compares one simulant to one reference material. Steps 4 – 6 in this methodology can be repeated using additional reference materials; e.g., the first reference material was a lunar highlands soil, but another can be chosen from among the lunar mare soils. Thus, a simulant can have multiple sets of FoMs comparing it to various space materials to understand how well it replicates the measured properties of each.

7. The sets of FoMs for each simulant are evaluated to determine the best simulants for each type of application, and the recommendations are recorded in the Fit-To-Use Table, part of the Simulant Users Guide (Schrader, et al., 2010). One simulant may be the best for tests of drilling in lunar highlands soil, while another may be better for drilling in lunar mare soil, a third may be best for oxygen extraction tests in lunar highlands soil, etc. Simulants may be listed as "most recommended," "recommended," "recommended for highlands," "recommended with reservations," or "not recommended" in each application based on their FoMs.

In the asteroid community there is a growing need to develop shared simulants to gain the benefits and avoid the pitfalls discussed above. With no availability of high fidelity simulants, many lower-fidelity simulants are being used by necessity. Housen (1992) used a mixture of 50% basalt fragments, 24% fly ash, 20% iron grit, and 6% water to simulate asteroid regolith for cratering ejecta experiments. Fujiwara et al. (2000) reported "various kind of rocks, sand, and artificial materials like bricks" while Yano et al. (2002) used glass beads and lunar soil simulant to study asteroid sampling via projectile impact. Sears et al. (2002) studied the formation of smooth regolith ponds on asteroids by using Martian regolith simulant JSC Mars-1 and with mixtures of sand plus iron grains. Sandel et al. (2006) reported the use of "meteorite simulants" for impact experiments to study collisional disruption and resulting fragment distribution from asteroids. Izenberg and Barnouin-Jha (2006) used playground sand with embedded cobbles to simulate asteroid regolith to study how impacts affect the morphology and vertical layering of asteroids. Makabe and Yano (2008) used bidisperse glass beads (0.5 mm and 5 mm) to study projectile impact in the Hayabusa-2 mission for capturing samples. Güttler et al. (2012) also used spherical glass beads to study crater formation on asteroid surfaces. Barucci et al. (2012) used a lunar regolith simulant and "many simulants" to test an asteroid sampling mechanism. Durda and co-workers (Durda et al., 2012; Durda et al., 2013; Durda et al., 2014) studied the morphology of asteroid surfaces using lunar soil simulant JSC-1A, glass microspheres, and bread flour. Bernold (2013) performed asteroid mining and conveying experiments using lunar regolith simulants. Crane et al. (2013) used shavings from a steel bar to simulate a Tholen Type M asteroid regolith for thermal inertia tests. Murdoch et al. (2013) studied the strength properties of asteroid regolith in microgravity using spherical soda-lime glass beads. Backes et al. (2014) used floral foam and "a variety of simulants" both hard and soft to represent the surface of a comet. Not included in this list are the many spectroscopic studies of terrestrial analog materials that inform remote observation and calibrate spacecraft instruments, nor space weathering experiments. This survey shows that, at the present, the materials chosen for asteroid simulants are generally lower-fidelity and lack the commonality and standards that would promote comparing results and improving the simulants. It also shows there is a wide range of uses for asteroid simulants.

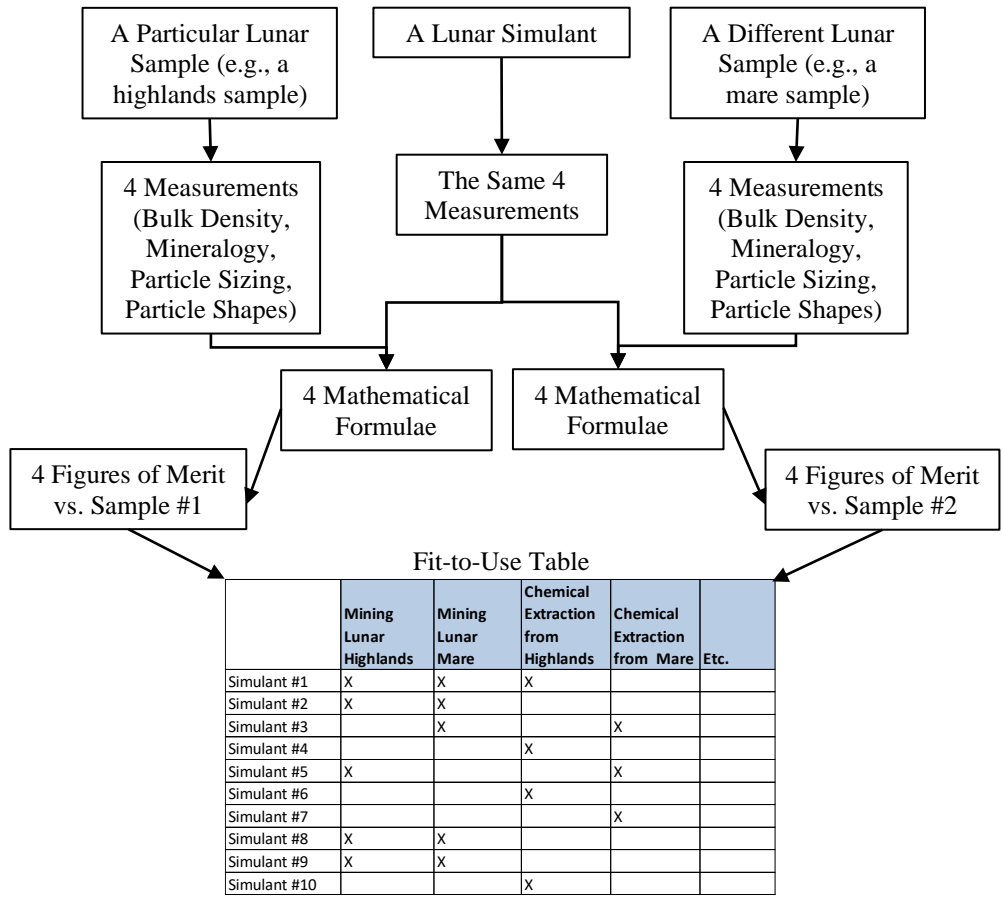

**Figure 1.** Summary of the NASA Simulants Team's methodology for determining applicability of lunar simulants to the identified uses.

In 2015, NASA awarded Deep Space Industries and the University of Central Florida a contract to develop asteroid simulants using the best practices developed in the NASA lunar simulants program. On October 6-7, 2015, the first asteroid simulants workshop (Metzger et al., 2016) was held at the University of Central Florida, co-sponsored by Deep Space Industries, the Florida Space Institute, and the Center for Lunar and Asteroid Surface Science (CLASS, a node of NASA's Solar System Exploration Research Virtual Institute, or SSERVI). The attendees reviewed the needs for asteroid simulant and decided that several asteroid classes would be produced, including CI, CM, C2, CV, L Ordinary, LL Ordinary, H Chondrite, Iron, Enstatite Chondrite, and Basaltic Chondrite types. Simulants would be based on the minerals found in meteorite types since mineralogy and crystal structure are a fundamental characteristic of rocks. The "root" and "branch" concept from the NASA lunar simulants program was adopted: the basic simulants would be developed as "roots" that the users can adapt into "branches" for their specialized purposes. For example, several carbonaceous chondrite meteorite types contain carcinogenic organic material, and a non-carcinogenic organic material will be used in the root simulants for ease in shipping and handling for most applications; health researchers who need the carcinogenic properties for their work can order the root simulant without the organic content and mix in their own carcinogenic organic matter at their facility. The workshop also decided the simulant would be available in several physical forms, and this plan has evolved so it is now four forms: the individual mineral components powdered and delivered in separate containers; a pre-mixed powder of all the minerals (all monomineralic grains); solid cobbles or slabs formed from the mixture of minerals bonded as a solid matrix; and regolith (formed by re-crushing the slabs) where the grains are now polymineralic in the appropriate mineral ratios. The workshop also identified 65 properties of asteroid materials that might be included in a simulant. Of these, the attendees chose 11 that would control the design of the simulants and another six that would be measured and reported but would not be control parameters. These properties are discussed below.

The workshop identified several of the meteorites that would be used for the reference materials: Orgueil for CI; Murchison for CM; Tagish Lake for C2; Allende for CV; Gibeon for Iron. One problem is that there is little pristine asteroid material available to guide development and validation of a simulant, since meteorites are just the part that survived entry (regolith did not survive) and the Hayabusa sample may not be representative of bulk regolith on a wide variety of asteroid sizes. This is in contrast to the lunar case where the Apollo and Lunakhod missions brought back many samples from many locations with corresponding geological context. To make up for this shortcoming, validation of the simulant will need to include other data sources in addition to tests upon the reference materials. The workshop attendees identified the following data sources: laboratory analysis of meteorites, bolide observations (to constrain compressive strengths), ground-based observations by radar and thermal infrared, observation of disrupted asteroids, spacecraft imagery, returned samples as they become available (Hayabusa, OSIRIS-REx, etc.), and modeling (interparticle cohesion, depletion of fines and buildup of lag surfaces, particle size distribution to match remote sensing, and theories of formation). The FoM formulas NASA developed for the lunar case require modification to include these data sources.

Lunar simulant had four FoMs developed: Bulk Density, Composition, Particle Sizing, Particle Shapes (Schrader et al., 2010). We have selected for initial development the following eight: Mineralogy, Elemental Composition, Average Grain Density, Bulk Density of Cobbles, Mechanical Strength of Cobbles, Magnetic Susceptibility, Volatile Release Patterns, and Regolith Particle Sizing. Section II will explain why these FoMs were selected and why several others were not selected. Sections III through X will discuss the eight FoMs in the order just listed. Section XI will be a discussion of the results, and Section XII will give the conclusions.

## II. Overall Methodology

The reference meteorites selected for each asteroid spectral class simulant are listed in Table 1. In this paper we present the only the CI simulant as proof-of-concept for the FoM methodology. FoMs are designed to validate a simulant in ways that are relevant to the proposed uses of the simulant (Rickman and Schrader, 2010). For asteroid simulant, some of the proposed uses are listed in Table 2, along with the relevant FoMs to validate it for those uses. There are a few notable quantities we have not included as FoMs. First, the angle of repose, internal friction, and cohesion are not adequately known from in situ asteroid regolith. We plan to measure and report these for each regolith simulant but we cannot quantify FoMs to evaluate them. Also, NASA considered these to be derivative properties resulting from the more fundamental particle size distribution, particle shapes, and bulk density, so they developed only FoMs for the fundamental properties. Second, to characterize the organic matter in carbonaceous simulant, we will use C-to-H ratio or the aliphatic/aromatic ratio or other metrics to compare simulants to meteorites. These have been evaluated for meteorites so we could develop an FoM for the organics. However, this is probably important for a more limited set of users so we leave an organic matter FoM to future development. Third, NASA's lunar simulants team defined a Particle Shapes FoM for lunar simulants, but they had much more data available on lunar particle shapes than we have for asteroid particle shapes so we leave this to future work. See Rickman, Immer, Metzger, et al. (2012) for the scale of effort in measuring a large sample of simulant particle shapes. Fourth, we will measure and report reflectance spectra for the simulants, but we do not believe it would be meaningful to reduce the spectra to a single number, as in an FoM.

It is an interesting mathematics question how to quantify the similarity of two geological materials, i.e., how near the materials are to each other in some "geological space". In measuring distances in ordinary space, such as the distance between opposite corners of a three-dimensional rectangular box, the Cartesian coordinates of a corner are the three elements of its position vector, and the difference between the opposite corners' position vectors is the diagonal vector $\vec{V}$, which has length

$$\|\vec{V}\|_n = \sqrt[n]{\Delta x^n + \Delta y^n + \Delta z^n}, \text{ with } n = 2 \quad (1)$$

This is the $\ell_2$ norm, where the subscript "2" indicates the exponent was $n = 2$. The geological space, on the other hand, is an $\ell_1$-normed space because we must use $n = 1$. The modal mineralogy of a geological material, its elemental composition, its particle size distribution, and possibly more features, can be considered vectors in $\ell_1$ because they are also lists of numbers that can be added and subtracted as geological materials are combined or compared. The elements of these lists are coordinates in the geological space. The totality of the material in regard to these quantities is simply the sum of the individual components. For example, all the mineralogical weight percentages in a rock add to 100% using $n = 1$ in Eq. 1. Since the sum is 100% (unity) by definition, it is a unit vector in $\ell_1$, indicating the "direction" of the rock in that space. Every rock is represented by another unit vector, so $\ell_1$ vector mathematics quantifies the differences between them. A generalization of the vector space is the Hilbert space where continuous functions may be considered infinite-dimensional vectors, each $(x, y)$ point along the curve being the coordinates in one of the (infinite) dimensions. The $\ell_1$ "length" or norm of a function is simply its integral, the area under the curve. An example is the integral of the differential particle size distribution, which is unity, so it is a unit function in $\ell_1$. The "distance" between two functions in $\ell_1$ is the integral of (the absolute value of) their difference, i.e., the absolute area between the curves. NASA initiated the use of $\ell_1$ norms for the FoM system. Here we extend the method and propose some improvements. We attempted to use generalized inner products or norms in different spaces but settled on the ones presented here as the most physically meaningful.

**Table 1.** Simulants and Reference Materials

| Simulant (Asteroid Class) | Selected Reference Material (Meteorite) |
|---|---|
| CI | Orgueil |
| CM | Murchison |
| C2 | Tagish Lake |
| CR | Average of five Antarctic finds: GRA 95229, LAP 02342, QUE 99177, PCA 91082, and GRA 06100 |
| CV | Allende |

**Table 2.** Some Uses for Asteroid Simulant and the Corresponding Figures of Merit

| Some Uses of Simulant | Figures of Merit | | | | | | | |
|---|---|---|---|---|---|---|---|---|
| | Mineralogical | Elemental Composition | Regolith Bulk Density | Cobble Density | Cobble Mechanical | Magnetic Susceptibility | Volatile Release Pattern | Particle Size Distribution |
| Technology development tests: | | | | | | | | |
| • Asteroid mining by mechanical methods | X | | X | X | X | | | X |
| • Asteroid mining by magnetic methods | X | X | X | X | | X | | X |
| • Extracting volatiles by thermal methods | X | | X | X | | | X | X |
| • Beneficiating mined asteroid materials | X | X | X | X | X | X | | X |
| • Chemically processing resources from beneficiated asteroid materials | X | X | | | | | X | X |
| • Planetary defense/asteroid redirection techniques | | | X | X | X | | X | X |
| • Radiation protection using asteroid mass | | X | X | X | | | | |
| • Anchoring methods | | X | X | X | X | X | | X |
| Pre-mission test of spacecraft hardware | | | | | | | | |
| • Thruster plume interactions with an asteroid | | | X | X | | | | X |
| • Landing gear impact on asteroid regolith | | | X | X | X | | | X |
| • Sample collection device | X | X | X | X | X | X | X | X |
| Astronaut health studies | | | | | | | | |
| • Exposure to dust | X | | | | | | | X |
| • Exposure to organic matter | X | | | | | | X | |
| Scientific Studies | | | | | | | | |
| • Space weathering | X | X | | | | | | X |
| • Impact dynamics | | | X | X | X | | | X |
| • Thermal characteristics of asteroids | X | | X | X | | | X | X |
| • Spectral characteristics of asteroids | X | | | | | | | X |

### III. Mineralogical Figure of Merit

For the Mineralogical FoM, $\Phi_M$, where the subscript indicates "mineralogical", we will use the method that was developed by NASA's lunar simulants team. NASA measured the modal mineralogy of the reference material and the simulant and wrote a vector for each,

$$\vec{R}_M = (r_1, r_2, r_3, \ldots, r_{N_M})^T$$
$$\vec{S}_M = (s_1, s_2, s_3, \ldots, s_{N_M})^T \quad (2)$$

where $\vec{R}_M$ is the reference material, $\vec{S}_M$ is the simulant, $r_i$ and $s_i$ are the weight percentages of the minerals in each, and $N_M$ is the total number of relevant minerals among both lists. The sum of the constituents must add to 100% (i.e., 1), so these are unit vectors under the $\ell_1$ norm (or taxicab norm),

$$\|\vec{R}_M\|_1 = \sum_{i=1}^{N_M} r_i = 1, \text{ and } \|\vec{S}_M\|_1 = \sum_{i=1}^{N_M} s_i = 1, \quad (3)$$

in "mineral space", which is an $\ell_1$–normed vector space, i.e., a Lebesgue space. $\Phi_M$ is defined as,

$$\Phi_M(\vec{S}_M, \vec{R}_M) = \sum_{i=1}^{N_M} \min(s_i, r_i) = \|\vec{S}_M \cap \vec{R}_M\|_1 \quad (4)$$

which defines the intersection operator ∩ in this context. It is the fraction of "overlap" in the compositions of the two materials, so $0 \leq \Phi_M \leq 1$.

For the mineralogical characterization of Orgueil we use the results of Bland et al. (2004) based on X-ray diffraction and Mössbauer spectroscopy. The methodology detects crystalline phases >0.5-1.0 wt% and iron-bearing phases >0.5 wt% but cannot detect amorphous, iron-poor phases, so there is a chance a small part of the composition was omitted. It is not economically viable to create a simulant with perfect fidelity, and the CI simulant we assess here was intended to include only phases >1 wt% so any phases omitted by Bland et al. would likely have been omitted from the simulant, anyway.

We found it necessary to re-bin several of the minerals in Orgueil. For the olivine forsterite-fayalite series, Bland et al. organized individual crystal measurements into five bins in the Mg-Fe series: Fo100, Fo90, Fo80, Fo60, Fo50 and Fo25, where FoX, x=X/100 represents (Mg$_x$, Fe$_{1-x}$)$_2$SiO$_4$. For the simulant we are concerned only with bulk chemistry, not spatial variability of the chemistry, so we re-bin these into bulk-equivalences of the end members,

$$w_{Fo100} = \sum_{\{X\}} \widehat{w}_{FoX} \left(\frac{X}{100}\right), \quad w_{Fo0} = \sum_{\{X\}} \widehat{w}_{FoX} \left(\frac{100-X}{100}\right) \quad (5)$$

where $\widehat{w}_{FoX}$ are the wt% values reported by Bland et al., and $w_{Fo100}$ and $w_{Fo0}$ are the equivalent forsterite and fayalite wt% values we use in calculating the FoM in Table 5, below.

We treat sulfur minerals similarly. Handling and shipping powdered pyrrhotite or troilite would be hazardous so we use only pyrite in the simulant, and we wish to grade its bulk chemistry for engineering test purposes. Orgueil's non-stoichiometric pyrrhotite (Fe$_{1-x}$S, 0<x<0.2, assuming

x=0.1 for calculations) is therefore converted into bulk-equivalences of the troilite (FeS) and pyrite (FeS$_2$),

$$w_{\text{FeS}} = \widehat{w}_{\text{FeS}} + \left(\frac{x}{1-x}\right)\widehat{w}_{\text{Fe}_{1-x}\text{S}}, \ w_{\text{FeS}_s} = \left(\frac{1-2x}{1-x}\right)\widehat{w}_{\text{Fe}_{1-x}\text{S}} \tag{6}$$

where $\widehat{w}_{\text{FeS}}$ and $\widehat{w}_{\text{Fe}_{1-x}\text{S}}$ are the wt% values reported by Bland et al., and $w_{\text{FeS}}$ and $w_{\text{FeS}_s}$ are the wt% values we use in the FoM in Table 5, below.

We also combine the phyllosilicates because we found their identification in modal analyses is insufficient for defining simulants. Bland et al. (2004), like others previously, identified Mg-rich serpentine plus a "disordered interstratified phase" of roughly equal amounts of saponite and serpentine. Serpentine could mean any of the four chrysotile polymorphs, antigorite, several other serpentine minerals of varied chemistry, or a mixture. Beck et al. (2014) argued from 10 µm band behavior that terrestrial Mg-rich serpentines are not good analogues for the highly disordered serpentine in CI and CM meteorites. As for the identification of saponite, in modal analyses it is often a catch-all to account for residual OH, Mg, Fe, and Si. (There is even some ambiguity of saponite chemistry in the literature; see, e.g., Anthony et al. 2001 vs. Wimpenny 2016.) Tomeoka and Buseck (1988) concluded it is saponite on the basis that it is not vermiculite, although the data were unable to distinguish saponite from chemically similar smectites. Bass (1971) identified montmorillonite (another smectite) instead of saponite. Calvin and King (1997) found that a linear mixing of 70% chamosite (a chlorite group phyllosilicate) with 30% antigorite (a serpentine) approximated the 5-25 µm spectrum of Orgueil reasonably well, especially over ~7-20 µm. Beck et al. (2010) report that any linear mixing of the spectra of five serpentine minerals and saponite fails to match the 10 µm band spectrum for the chondrites they analyzed including Orgueil, so they conclude the disordered phyllosilicates are distinct from terrestrial phyllosilicates. Beck et al. (2014) interpreted 10 µm features in Orgueil as "consistent with saponite like phyllosilicates" although they also found terrestrial saponites are distinct so they are not good analogues. King et al. (2015) followed prior authors who labeled the phyllosilicates as serpentine and saponite and did not attempt to deconvolute them further, using an x-ray standard consisting of a disordered serpentine/saponite mixture. We found no analysis that obtained a more specific identification of the phyllosilicates than these. These studies suggest that simply selecting terrestrial sources for a serpentine and a saponite may not sufficiently replicate the Orgueil's mineralogy to produce the correct chemical and volatile release behaviors, so further guidance is needed.

Our first attempts to create simulants found that water was thermally released at lower temperatures than the reference meteorites, so we decided to select phyllosilicates with guidance from their water release patterns while still choosing from the broad groups indicated by modal analyses. We also gave consideration for the strength of the resulting cobbles, since the phyllosilicates are the only binder in the CI simulant and we need them to produce realistically strong cobbles. For the mineralogical FoM, we binned the serpentine and saponite of the modal analyses together into the broader category "phyllosilicates"; more specific categories would be meaningless in the mineralogical FoM calculation. We rely upon the volatiles release FoM and cobble strength FoM for more specific measurement of their merit.

The organic content of CI carbonaceous chondrites is studied in considerable detail to identify organic species (e.g. Cody and Alexander, 2005), but is much less well studied for bulk organic content. These species are not susceptible to x-ray diffraction analysis and are not reported in the sources that focus on mineralogy (i.e. Bland et al., 2004). The available bulk data is largely from studies of elemental composition which report weight percentages of carbon (Wasson and Kallemeyn, 1988). Our assumption is that the bulk of the carbon is in the form of soluble and insoluble organic carbon compounds. These compounds will include nitrogen and hydrogen. We therefore added 5 wt% organic material to the simulant recipe and re-normalized the modal analysis. We chose sub-bituminous coal as simulant for the organic material in Orgueil because it avoids the most severe health hazards of polycyclic aromatic hydrocarbons in meteoritic organic material, making the simulant safe for a wider range of users, yet it matches the aromaticity and elemental composition of organics in Orgueil. Aromaticity for several meteorites (Cody and Alexander, 2005) and coal grades (Odeh, 2015) are compared in Table 3.

**Table 3.** Aromaticity for several meteorites and coal grades.

|  |  | Meteorite (type) | | | |
|---|---|---|---|---|---|
|  |  | Aromaticity (%) | | | |
|  |  | EET92042 (CR2) | Orguiel (CI1) | Murchison (CM2) | Tagish Lake (C2) |
| Coal Grade | Aromaticity (%) | 48 - 52 | 61 - 65 | 62 - 66 | 79 – 83 |
| Lignite | 49 | X |  |  |  |
| Sub-bituminous | 60-73 |  | X | X |  |
| Bituminous | 62-80 |  |  | X | X |
| Semi-anthracite | 85 |  |  |  |  |
| Anthracite | 91 |  |  |  |  |

Note. Meteoritic data: Cody and Alexander (2005). Coal data: Odeh (2015).

The recipe for the simulant UCF/DSI-CI-2 is shown in Table 4. To calculate the FoM we assumed the source materials are pure; X-ray diffraction analysis of these materials showed some minor impurities are present, for example dolomite in the antigorite, and a serpentine phase in the olivine. These impurities are unavoidable and may result in a slightly lower FoM than reported here, but quantifying these effects is outside the scope of developing and testing the FoM methodology. The olivine Fo90 is re-binned into equivalent mass fractions of the forsterite and fayalite. The re-binned mineralogy for both Orgueil and the CI simulant are shown in Table 5. The calculated mineralogical FoM is $\Phi_M = 0.83$, meaning that 83% of the simulant's composition "overlaps" the meteorite's composition in the defined material groupings.

**Table 4.** Recipe for CI Asteroid Simulant

| Mineral/Material | wt% |
|---|---|
| Antigorite $(Mg,Fe_{2+})_3Si_2O_5(OH)_4$ | 48.0 |
| Vermiculite $(Mg,Fe,Al)_3(Al,Si)_4O_{10}(OH)_2 \times 4(H_2O)$ | 9.0 |
| Attapulgite $(Ca,Na)_{0.33}(Mg_{2.66},Li_{0.33})Si_4O_{10}(F,OH)_2 \times 4H_2O$ | 5.0 |
| Olivine Fo90 $(Mg_{0.9},Fe_{0.1})_2SiO_4$ | 7.0 |
| Magnetite $Fe_3O_4$ | 13.5 |
| Pyrite $FeS_2$ | 6.5 |
| Epsomite $MgSO_4 \times 7H_2O$ | 6.0 |
| Sub Bituminous coal | 5.0 |
| **TOTAL** | **100.0** |

**Table 5.** Re-binned mineralogy of the Orgueil meteorite (Bland et al., 2004) and CI Simulant with FoM calculations.

| Mineral/Material | Orgueil mass fraction $r_i$ | Simulant mass fraction $s_i$ | FoM Calculation $\min(s_i, r_i)$ |
|---|---|---|---|
| Combined Phyllosilicates | 0.6793 | 0.6200 | 0.6200 |
| Equivalent Fayalite $FeSiO_4$ | 0.0120 | 0.0070 | 0.0070 |
| Equivalent Forsterite $MgSiO_4$ | 0.0564 | 0.0630 | 0.0564 |
| Magnetite $Fe_3O_4$ | 0.0922 | 0.1350 | 0.0922 |
| Equivalent FeS | 0.0580 | 0.0000 | 0.0000 |
| Equivalent $FeS_2$ | 0.0048 | 0.0650 | 0.0048 |
| Ferrihydrite $(Fe_{3+})_2O_3 \times 0.5H_2O$ | 0.0475 | 0.0000 | 0.0000 |
| Epsomite $MgSO_4 \times 7H_2O$ | 0.0000 | 0.0600 | 0.0000 |
| Organics | 0.0500 | 0.0500 | 0.0500 |
| **TOTAL** | 1.00 | 1.00 | $\mathbf{\Phi_M = 0.83}$ |

## IV. Elemental Figure of Merit

NASA's lunar simulants team did not define an Elemental FoM, $\Phi_E$ (where the E subscript indicates elemental), but we find it useful for asteroids. For example, in developing the CI1 simulant we attempted to match the mineralogy of the Orgueil meteorite, which has (re-binned equivalent) 5.8 wt% FeS. This is unstable in powder form so would be hazardous to process and ship in that form. Substituting anything for it will reduce $\Phi_M$ by 0.058 regardless what we substitute, so if $\Phi_M$ told the entire story there would be no guidance to choose among substitutes. However, using other sulfur compounds such as pyrite $FeS_2$ could increase the fidelity of the simulant for radiation shielding studies, since radiation does not care what crystal structure each element belongs to as long as the elements are present in correct proportion. By judicious choice of the substituted minerals, the overall elemental composition can be brought very close to that of the reference material without further reducing $\Phi_M$. We motivate this secondary selection by defining $\Phi_E$,

$$\vec{R}_E = (u_1, u_2, u_3, \ldots, u_{N_E})^T, \quad \|\vec{R}_E\|_1 = 1$$

$$\vec{S}_E = (v_1, v_2, v_3, \ldots, v_{N_E})^T, \quad \|\vec{S}_E\|_1 = 1 \tag{7}$$

$$\Phi_E(\vec{R}_E, \vec{S}_E) = \|\vec{S}_E \cap \vec{R}_E\|_1 \tag{8}$$

where $u_i$ and $v_i$ are the weight percentages of each element, and $N_E$ is the number of relevant elements. This "element space" is another $\ell_1$-normed vector space. An isotopic FoM could also be defined but at present we cannot justify the much higher expense in developing an isotopic simulant for the additional fidelity it would bring.

For elemental decomposition of Orgueil we use the modal mineralogy of Bland et al. (2004) summing elements per the mineral formulas to find $\vec{R}_E$ and $\vec{S}_E$. For phyllosilicates Bland et al. used the compositions from Tomeoka and Buseck (1988) for a Fe-bearing, Mg-rich serpentine and a saponite-serpentine disordered interstratified phase. Regardless whether those phyllosilicate identifications are accurate, the modal decomposition was based on the elemental abundances per those formulas so reconstructing elemental abundances with the same formulas is correct. For the organics in Orgueil and the sub bituminous coal in the simulant we used $C_{100}H_{77.4}S_{0.5}N_{1.1}O_{15.9}$ calculated for sub-bituminous coal from Zumdahl and Zumdahl (2009) in general agreement with Odeh (2015), but it neglects the ash content. The results for both Orgueil and simulant are given in Table 6. We calculate $\Phi_E = 0.94$.

**Table 6.** Elemental FoM Calculation

| Element | Orgueil mass fraction $u_i$ | Simulant mass fraction $v_i$ | FoM Calculation $\min(u_i, v_i)$ |
|---|---|---|---|
| Fe | 0.1895 | 0.1624 | 0.1624 |
| Si | 0.1064 | 0.1118 | 0.1064 |
| Mg | 0.0962 | 0.1354 | 0.0962 |
| S | 0.0525 | 0.0419 | 0.0419 |
| C | 0.0322 | 0.0385 | 0.0322 |
| H | 0.0202 | 0.0167 | 0.0167 |
| Al | 0.0065 | 0.0114 | 0.0065 |
| Ni | 0.0100 | 0.0015 | 0.0015 |
| Ca | 0.0087 | 0.0150 | 0.0087 |
| Na | 0.0055 | 0.0004 | 0.0004 |
| N | 0.0012 | 0.0005 | 0.0005 |
| Cr | 0.0024 | 0.0003 | 0.0003 |
| Mn | 0.0017 | 0.0003 | 0.0003 |
| P | 0.0013 | 0.0004 | 0.0004 |
| O and traces | 0.4662 | 0.4634 | 0.4634 |
| **Total** | 1.0000 | 1.0000 | $\mathbf{\Phi_E = 0.94}$ |

## V. Average Grain Density Figure of Merit

We develop a density FoM, $\Phi_D$, for the simulant. The density FoM was defined for lunar regolith simulant by NASA (Hoelzer, et al., 2011) as three separate measurements including the average mineral density of the assemblage of mineral grains $\rho_g$, the minimally compacted bulk density of the regolith $\rho_{\min}$ (corresponding to 0% relative density [Carrier et al., 1991]), and the maximally compacted bulk density $\rho_{\max}$ (100% relative density). For asteroid regolith, we do not have adequate data to know $\rho_{\max}^R$ or $\rho_{\min}^R$, so we restrict our density FoM to only the average grain density. This grain density applies to both regolith and cobble versions of the simulant, although for cobbles we do have porosity information from the meteorites so will also calculate a cobble bulk density FoM in the next section. The average mineral density is defined (here with superscripts $R$ for reference material, alternatively $S$ for simulant), as

$$\rho_g^R = \sum_{i=1}^{N_M} r_i \rho_i, \tag{9}$$

Where $r_i$ is the mass fraction of each mineral in the reference meteorite (or $s_i$ for the simulant), and $\rho_i$ is the published mineral density for each mineral. Hoelzer et al. (2011) defined the lunar regolith FoM for average grain density as,

$$\Phi_D = \max\left\{0, 1 - \frac{1}{w}\frac{|\rho_g^S - \rho_g^R|}{\rho_g^R}\right\} \tag{10}$$

so $\Phi_D = 1$ when $\rho_g^S = \rho_g^R$ but $\Phi_D$ diminishes linearly to zero as $\rho_g^S \to \rho_g^R(1 \pm w)$, and $\Phi_D = 0$ when $|\rho_g^S - \rho_g^R| > (1 + w)$ as shown in Fig. 2. The scale factor $0 < w < 1$ is a number chosen somewhat arbitrarily to indicate how far the simulant's density can be from the reference material's density before an engineering test using this simulant has "no value". NASA did not publish a selected value of $w$ but the geologist who led that project, D. Rickman (personal communication, 11/13/2017), suggests using $w = 0.5$, which we follow here. This indicates that a simulant more than 50% off the correct density value has "no value" for mechanical testing purposes, whereas the testing value increases linearly as the density error reduces to 0%.

This method, where expert opinion selects the value of the scale factor $w$, is not mathematically rigorous, and different values of $w$ must be selected for the different FoMs, below. This may cause the system to seem arbitrary. However, this is the method NASA developed, and we have not identified a better one. Theoretically predicting from first-principles the quantitative value of a test that involves geological materials is not currently possible. Perhaps more theoretical progress will improve the situation. For now, we recommend this system to the community as the best practical method.

For the reference value of Orgueil we relied upon the measurements by Consolmagno and Britt (1998). They found $\rho_g^R = 2.43 \pm 0.06$. We measured grain density of the simulant and found $\rho_g^S = 2.74 \pm 0.01$. We suspect the value is high because the organic content is a little low in the simulant, causing the denser constituents to be a little high. We calculate $\Phi_D = 0.75$. These relationships are shown in Fig. 2.

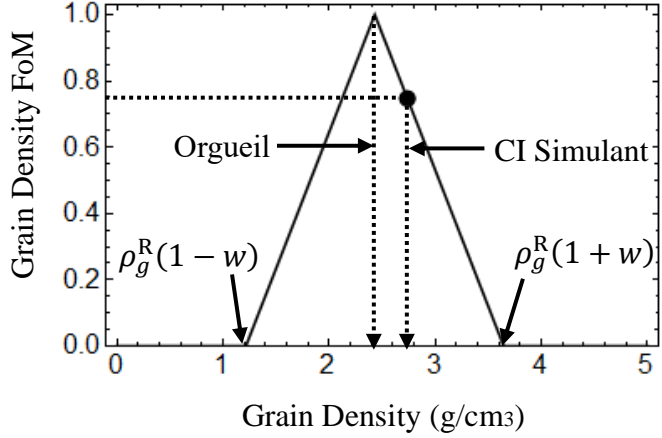

**Figure 2.** Grain Density FoM $\Phi_D$ vs. Grain Density (g/cm3) per Eq. 10 with $\rho_g^R = 2.43$ g/cm3 and $w = 0.5$. The solid circle is $\Phi_D = 0.75$ at $\rho_g^S = 2.74$ g/cm3.

## VI. Cobble Bulk Density Figure of Merit

For the cobble version of the simulant we also develop an FoM for bulk density, $\Phi_{BD}$. Because cobbles have porosity $n$, $\rho_{bulk}^R = \rho_g^R(1 - n^R)$ and $\rho_{bulk}^S = \rho_g^S(1 - n^S)$, so we define the Bulk Density FoM,

$$\Phi_{BD} = \max\left\{0, 1 - \frac{1}{w}\frac{|\rho_{bulk}^S - \rho_{bulk}^R|}{\rho_{bulk}^R}\right\} \quad (11)$$

again with $w = 0.5$. We report only $\Phi_D$ for regolith versions of asteroid simulant but we report both $\Phi_D$ and $\Phi_{BD}$ for cobble versions. (The value of $\Phi_D$ is the same for both since the regolith is made from crushed cobbles.)

For the reference value of Orgueil we rely upon the measurements by Consolmagno and Britt (1998). They found $\rho_{bulk}^R = 1.58 \pm 0.03$, which indicates porosity of ~35%. We measured the simulant and found $\rho_{bulk}^R = 1.80 \pm 0.01$, which indicates porosity of ~34%. We note the porosities are consistent, so the difference in bulk densities is due to the difference in grain densities, probably due to lower organic content. We calculate $\Phi_{BD} = 0.72$. These relationships are shown in Fig. 3.

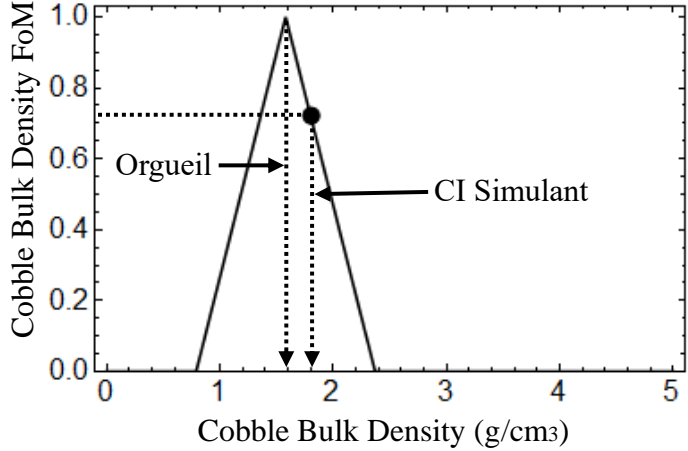

**Figure 3.** Grain Density FoM $\Phi_{BD}$ vs. Grain Density (g/cm3) per Eq. 11 with $\rho_{bulk}^{R} = 1.58$ g/cm3 and $w = 0.5$. The solid circle is $\Phi_D = 0.72$ at $\rho_{bulk}^{S} = 1.80$ g/cm3.

### VII. Cobble Mechanical Strength Figure of Merit

The lunar simulants evaluated by NASA included only regolith simulants, but here we are evaluating both regolith and cobble versions of asteroid simulant. Therefore, we also define a cobble strength FoM, $\Phi_{CS}$, using the unconfined compression strength $U$ of the material. Effective strength of a meteorite is commonly believed to decrease as the size increases according to Weibull's statistical theory (Weibull, 1951),

$$\sigma = \sigma_S (m_S/m)^\alpha \qquad (12)$$

where $\sigma$ is effective strength of the object, $\sigma_S$ is strength for a smaller-sized cobble, $m$ is mass of the object, $m_S$ is mass of the smaller-sized cobble, and the index $\alpha$ has been estimated over a wide range from 0.03 to 0.5 depending on the rock material or other conditions (Yoshinaka et al., 2008; Svetsov et al., 1995; Popova, et al., 2010). Equivalently, $\sigma = \sigma_S (V_S/V)^\alpha$ using volumes instead of mass.

The asteroid simulant workshop in 2015 identified bolide observations and laboratory measurements of meteorites as the two reference data sets for $\Phi_{CS}$ (Metzger et al., 2016). An extensive review of meteoroid fragmentation in the atmosphere compared to meteorite strengths finds that "there is almost no correlation between apparent strength and meteoroid mass," and the very low strengths of the larger bodies in flight must be due to a highly fractured state that does not match expectations of the Weibull law (Popova, et al., 2011). Therefore, we cannot reliably use bolide or other atmospheric flight observations as reference measurements to evaluate merely cobble-sized simulants. This leaves laboratory measurements of meteorites. Significant scatter can be expected in these results, too. Hogan et al. (2015) identified two different fracture mechanisms: "one associated with the structural failure of material and one associated with inherent microstructure length scales (i.e., size and spacing of defects)," and while some rock types failed as a function of strain rate, other rock types did not (Hogan et al., 2015). Yamaguchi (1970) found that ten or more rock samples must be measured to obtain a statistically meaningful result, "even if all the test-pieces are prepared from the same block of the rock." It is likely most measured meteorite strengths are based on less than ten samples. For this reason, we adopt a *reference model* for meteorite strengths derived from analysis of the existing sparse data sets. This model will need configuration control and version numbers as better data become available to effectively communicate pedigree of the simulants.

Rock strengths vary over orders of magnitude, so NASA's FoM method for scalars like density should be adapted to a logarithmic version,

$$\Phi_{CS} = \max\left\{0, 1 - \frac{|\log_{10} U^S - \log_{10} U^R|}{\log_{10} w}\right\} = \max\left\{0, 1 - \frac{|\log_{10}(U^S/U^R)|}{\log_{10} w}\right\} \quad (13)$$

where $U^S$ and $U^R$ are unconfined compressive strength for the simulant material and reference model, respectively, and $w = 5$ means a test with the simulant has "no value" if the simulant is stronger or weaker by more than a factor of 5. As with the previous FoMs, this is NASA's method to use the judgement of expert opinion, and we recommend it as a practical choice. Members of the community are invited to identify a better method or to argue that a different value of $w$ is more appropriate in any of the FoMs to improve the system.

For mechanical strength of cobbles, a literature search failed to find any measurement of the compressive or tensile strength of cobble-sized or gravel-sized fragments of Orgueil. Tsuchiyama et al. (2008) measured fragments 50-200 µm in size that were created by gently crushing meteoritic samples. In six micro-indenter measurements, Orgueil averaged 3.1 MPa tensile strength. They estimated, by scaling from Weibull (1948) theory supported with some empirical data, that a fragment 10 cm across would have tensile strength about 0.2 MPa. A 4.4 cm fragment scales to 0.43 MPa by that relationship. Popova et al. (2011) and references therein report the compressive and/or tensile strengths of meteorites including seven that have both. The seven ratios of compressive to tensile strength average 7.7 ± 1.5. This predicts 2.5 MPa compressive strength for a 4.4 cm fragment of Orgueil. Error bars are difficult to estimate due to uncertainty in the Weibull scaling, which attempts to quantify the size-dependent weakening of a material through propagation of cracks and internal flaws and therefore requires a large sample of measurements of many sizes. This is not possible in the limited meteorite collection. However, it is still highly valuable to communicate the strengths of simulants, so this will serve as a reference model until better data are available. Version 1.0 of the reference model for mechanical strength of cobbles is,

$$U^R = 0.77 \left(\frac{V}{1 \text{ m}^3}\right)^{1/8} \text{ MPa} \quad (14)$$

where *V* is the volume of the reference sample.

For the Cobble Mechanical Strength FoM, we measured the compressive strength of a cobble of CI simulant formed in a cylindrical mold with post-drying dimensions 38.173 mm diameter × 76.346 mm high. We followed testing standard ASTM C39/C39M-17b using an MTS Criterion

Model 43 tester. Four specimens were tested at a loading rate of 3 mm/s with a 5 kN load cell. The cobble failed at an average of 1.7 ± 0.5 MPa. The volume of this sample was the same as a cube with side 4.4 cm. By Eq. 14, the reference model for 4.4 cm scales to 2.5 MPa. By Eq. 13 with $U^S = 1.7$ MPa and $U^R = 2.5$ MPa we calculate $\Phi_{CS} = 0.77$. Fig. 4 is a lin-log plot of Eq. 13 showing these relationships.

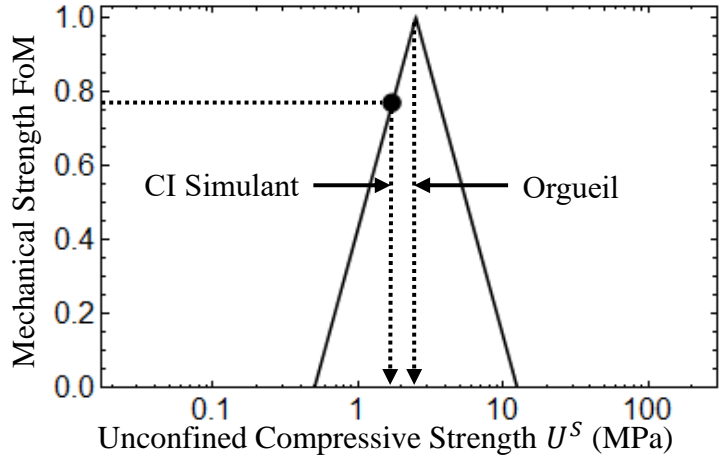

**Figure 4.** Log-linear plot of Cobble Mechanical Strength FoM $\Phi_{CS}$ vs. Unconfined Compressive Strength (MPa) per Eq. 13 with $U^R = 2.5$ MPa and $w = 5$. The solid circle is $\Phi_{CS} = 0.77$ at $U^S = 1.7$ MPa

### VIII. Magnetic Susceptibility Figure of Merit

The magnetic properties of asteroid simulant are important for testing technologies that use magnetism to manipulate asteroid material, including anchoring a spacecraft, mining, and beneficiating mined material. A not-insignificant fraction of the susceptibility may be provided by the phyllosilicates, which are inadequately identified in Orgueil, so we develop a magnetic susceptibility FoM to augment the mineralogical FoM. Carbonaceous meteorites may have paramagnetic, ferromagnetic, and superparamagnetic contributions to the apparent susceptibility but a full thermomagnetic characterization is beyond the scope of this first effort. Instead, we measure magnetic susceptibility $\chi$ of simulants at room temperature and compare to similar measurements of meteorites. Because susceptibility varies over orders of magnitude define a logarithmic FoM similar to the compressive strength FoM:

$$\Phi_{MS} = \max\left\{0, 1 - \frac{|\log_{10}(\chi^S) - \log_{10}(\chi^R)|}{\log_{10} w}\right\} = \max\left\{0, 1 - \frac{|\log_{10}(\chi^S/\chi^R)|}{\log_{10} w}\right\} \quad (15)$$

where $\chi$ is in 10-9 m3/kg, and again we choose $w = 5$. For magnetic susceptibility of Orgueil, measurements are listed in Table 7. A mass-weighted average of the three rows yields our reference value $\chi^R = 59{,}906$ 10-9 m3/kg, or $\log_{10} \chi^R = 4.78$. We used a Faraday Scale to measure the susceptibility of the CI simulant and we obtained $\chi^S = 63{,}850$ 10-9 m3/kg or $\log_{10} \chi^S = 4.81$, which is in the range of Orgueil measurements. We calculate $\Phi_{MS} = 0.96$. These relationships are shown in Fig. 5.

**Table 7.** Orgueil Magnetic Susceptibility Measurements

| Sample | Fall? | Mass (g) | $\log_{10} \chi$ in $10^{-9}$ m³/kg | $\chi$ ($10^{-9}$ m³/kg) | Reference |
|---|---|---|---|---|---|
| Vatican 719 | Fall | 14.0 | $4.11 \pm 0.08$ | 12,822 | Macke, Consolmagno, and Britt, 2011 |
| Vatican 718 | Fall | 47.2 | $4.86 \pm 0.08$ | 72,444 | Macke, Consolmagno, and Britt, 2011 |
| Combination of 9 samples | | 190.4 | $4.78 \pm 0.05$ | 60,256 | Rochette et al. 2008 |

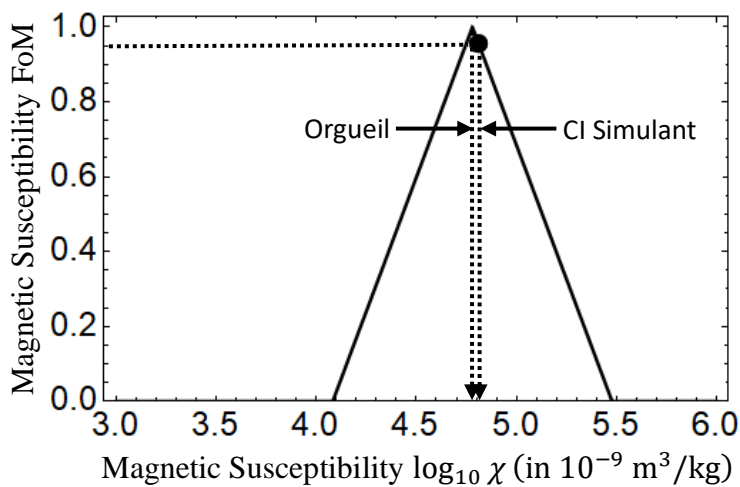

**Figure 5.** Plot of Magnetic Susceptibility FoM $\Phi_{MS}$ vs. Magnetic Susceptibility $\log_{10} \chi$ (in $10^{-9}$ m³/kg) per Eq. 15 with $\log_{10} \chi^R = 4.78$ and $w = 5$. The solid circle is $\Phi_{MS} = 0.96$ at $\log_{10} \chi^S = 4.81$.

## IX.   Volatile Release Figure of Merit

We define a Volatile Release FoM, $\Phi_{VR}$, based upon differential thermal gravimetry (DTG) at ambient pressure. The volatile release pattern depends not only on the mineralogy but also on the initial volatile inventory at the time DTG was performed. This initial volatile inventory depends on (1) the hydration and hydroxylation state of the source minerals that went into the simulant, (2) the method that the simulant was wetted and dried to make cobbles (prior to recrushing to make regolith), and (3) the subsequent handling and storage including exposure to humidity, heat and vacuum. Here we used freshly manufactured simulant to avoid the latter. The mineralogical analyses of meteorites in the literature leave some ambiguity about the specific phyllosilicates. This left some freedom to the simulant developers, independent of the resulting $\Phi_M$, to choose phyllosilicates that either improve or do not improve how well the DTG curve matches that of a reference material. The fidelity of DTG curves is important for mining technology tests of volatile extraction and scientific studies of asteroid volatiles. The differential volatile release curves of the simulant and reference material are $\Omega_S = \Omega_S(T)$ and $\Omega_R = \Omega_R(T)$ in wt%/°C. The total weight percentages of the samples that are released as volatiles over the full temperature range are the $\ell_1$-norms of the DTG curves,

$$V_S = \|\Omega_S\|_1 = \int_{T_{\min}}^{T_{\max}} |\Omega_S|\, dT, \quad V_R = \|\Omega_R\|_1 = \int_{T_{\min}}^{T_{\max}} |\Omega_R|\, dT \tag{16}$$

where $T_{\min}$ and $T_{\max}$ define the temperature range used in DTG per standard practice, typically ambient to 1000 °C. The absolute value symbols may be dropped since the volatile release curves are everywhere positive. $\Omega_S$ and $\Omega_R$ are not unit functions in the $\ell_1$-normed Lebesgue space because $V_S < 1$ and $V_R < 1$. These norms and the patterns of the temperatures at which volatiles are released (i.e., the magnitude and direction of $\Omega$ in $\ell_1$ space) are independent to some degree and both vital to the FoM so instead of performing a (pseudo) inner product as we have for several other FoMs, we may wish to base $\Phi_{VR}$ on the norm of the difference function,

$$\delta\Omega = \|\Omega_S - \Omega_R\|_1 = \int_{T_{\min}}^{T_{\max}} |\Omega_S - \Omega_R|\, dT. \tag{17}$$

scaling it as we did the density FoMs,

$$\Phi_{VR}^{(1)} = \max\left\{0, 1 - \frac{1}{w}\frac{\delta\Omega}{V_R}\right\}. \tag{18}$$

where the superscript indicates that this was the first version of the FoM. Examining this formulation, we identify a problem: slight shifts in the release temperatures could create large reductions in the FoM, even though the same amount of volatiles were released at almost the same temperatures. This motivated us to use cumulative volatile release functions instead,

$$v_S(T) = \int_{T_{\min}}^{T} |\Omega_S(T')|\, dT', \quad v_R(T) = \int_{T_{\min}}^{T} |\Omega_R(T')|\, dT',$$

$$\|v_R\|_1 = \int_{T_{\min}}^{T_{\max}} |v_R|\, dT, \quad \delta v = \|v_S - v_R\|_1 = \int_{T_{\min}}^{T_{\max}} |v_S - v_R|\, dT$$

$$\Phi_{VR} = \max\left\{0, 1 - \frac{1}{w}\frac{\delta v}{\|v_R\|_1}\right\}. \tag{19}$$

This penalizes the simulant if it releases too much or too little of the volatiles by each temperature, but because it is an integral of the cumulative difference the penalty is limited by how quickly the two cumulative release curves come together again. Thus, if it releases the right amount of volatiles at just a slight deviation in temperature, the penalty is slight. This is non-dimensionalized by normalizing with the total area under the reference curve, $\|v_R\|_1$.

For volatile release characterization of Orgueil, we rely upon DTG of two samples performed by King et al. (2015). They heated each sample at 10 °C/min from ~25 - 990 °C. We measured the volatile release patterns of the CI simulant at the Kennedy Space Center on a TA Instruments Simultaneous DSC/TGA Q600. The resulting DTG curve and the cumulative volatile release curve are shown in Fig. 6 along with the average of the two measurements of Orgueil by King et al. (2015). The simulant successfully replicated the major pattern shapes (except it has a release feature of unknown cause between 448.7 °C to 450.6 °C that does not exist in Orgueil, releasing only a small quantity of volatiles as seen in Fig. 6, right, because it is so narrow). However, by 991 °C the simulant released only 19.8 wt% compared to 30.5 ± 1.5 wt% for Orgueil. With $w = 0.5$, we calculated $\Phi_{VR} = 0.42$ and with $w = 1$ we calculated $\Phi_{VR} = 0.53$. The last of these two figures seems appropriate, since the simulant's net released volatiles was 65% of intended and the patterning was good so it should not be de-rated too much below 0.65. We therefore choose $w = 1$.

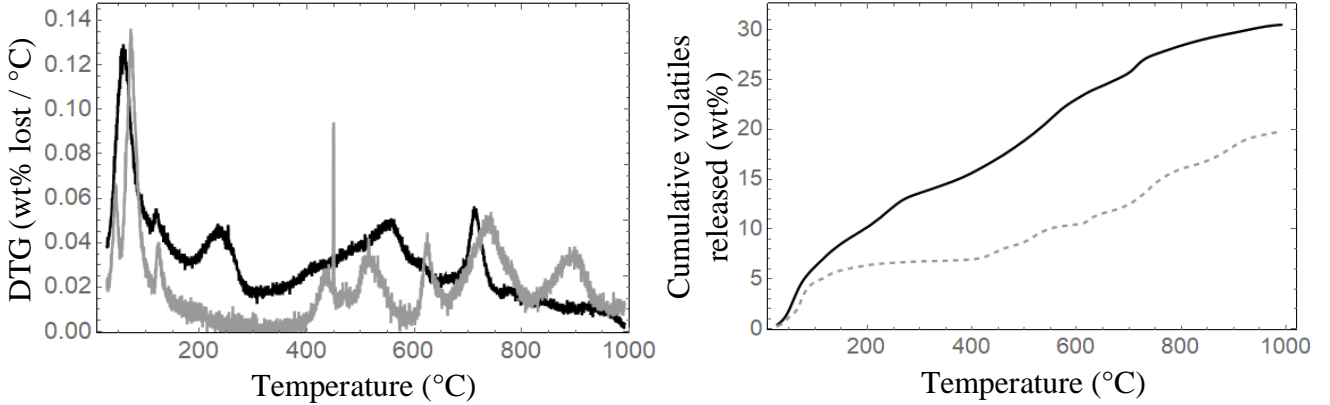

**Figure 6.** (Left) Differential Thermo-Gravimetric curves: Black, $\Omega_R(T)$ for Orgueil from King et al. (2015); Gray, $\Omega_S(T)$ for the CI simulant. (Right) Cumulative volatile release: Black, $v_R(T)$ for Orgueil from King et al. (2015); Gray dashed, $v_S(T)$ for the CI simulant.

## X. Regolith Particle Size Figure of Merit

The only returned sample of asteroid regolith so far was by Hayabusa from a smooth terrain on Itokawa (Tsuchiyama et al., 2011). We have no samples from the differently sized asteroids, different types of terrain, or different depths beneath the surface, so our data set is not sufficiently diverse for any level of confidence. In contrast, because of the large number of returned samples from the Moon, we have confidence in our understanding of lunar regolith particle size distributions including the average and extremal cases. Metzger and Britt (2018) examined the available data sets for asteroid regolith particle sizing and developed a reference model in lieu of a reference sample against which to calculate the FoM of a simulant. Version 1.0 of the Particle Size Distribution Reference Model is the following differential particle number distribution,

$$n_R(D) = \begin{cases} c_1 D^{-2.5}, & D_{Min}^{Surface} \geq D \geq D_{Max}^{Surface}, & \text{for surface deposits} \\ c_2 D^{-3.5}, & D_{Min}^{Bulk} \geq D \geq D_{Max}^{Bulk}, & \text{for bulk regolith} \end{cases}$$

$$1/c_1 = \int_{D_{Min}^{Surface}}^{D_{Max}^{Surface}} D^{-2.5} dD, \quad 1/c_2 = \int_{D_{Min}^{Bulk}}^{D_{Max}^{Bulk}} D^{-3.5} dD \qquad (20)$$

where $D_{\text{Min}}^{\text{Surface}}$, $D_{\text{Max}}^{\text{Surface}}$, $D_{\text{Min}}^{\text{Bulk}}$, and $D_{\text{Max}}^{\text{Bulk}}$ are user-selectable parameters to represent the asteroid they wish to simulate according to their best available model, since data indicate these parameters may depend on size and history of the body, specific terrain, or other factors. This can be converted to the volume-weighted differential particle size distribution,

$$f_R(D) = \frac{n_R(D)D^3}{\int_{D_{\min}}^{D_{\max}} n_R(D')D'^3 dD'} \tag{21}$$

Assuming all particle size ranges have the same mineralogical composition and microporosity, this is equivalent to the mass-weighted differential particle size distribution. This may also be converted to the cumulative mass-finer-than distribution,

$$F^R_\leq(D) = \int_{D_{\min}}^{D} f_R(D')\,dD' \tag{22}$$

$$F^R_\leq(D) = \begin{cases} \max\left(0,\ \min\left(1,\ \frac{D^{4+q}-D_{\min}^{4+q}}{D_{\max}^{4+q}-D_{\min}^{4+q}}\right)\right), & \text{if } q \neq 3 \\ \max(0,\ \min(1,\ D/D_{\max})), & \text{if } q = 3 \end{cases} \tag{23}$$

We considered three methods to calculate a FoM based on this reference model.

Method 1

The NASA lunar simulants team defined the Particle Size Distribution FoM,

$$\Phi_{\text{PSD}} = \int_0^\infty \min(f_S, f_R)\,dD \tag{24}$$

where $f_S = f_S(D)$ is the simulant's volume- (mass-) weighted, differential particle size distribution. $f_R$ and $f_S$ are unit-normalized functions in an $\ell_1$ Lebesgue space. Particle sizing is often determined by sieving, which discretizes the continuous function into $N$ size ranges per the available sieve screens, so the NASA lunar simulants team discretized this formulation,

$$\widehat{\Phi}_{\text{PSD}} = \sum_{i=1}^{N} \min(f_i^S, f_i^R) \tag{25}$$

where $f_i^S$ and $f_i^R$ are the fraction of a material's mass in the range of particle diameters $D_i \leq D < D_{i+1}$ as defined by the sieve screens. This introduces ambiguity in $\widehat{\Phi}_{\text{PSD}}$ because two sets of sieve screens will generally produce two different values for $\widehat{\Phi}_{\text{PSD}}$. Here we attempt to eliminate the ambiguity by developing a continuous formulation. It is based on cumulative size distributions and it interpolates between measured data points to approximate a continuous function.

We identified another problem with this method. Consider the extreme case where $n_R$ is a mono-sized packing of particles of size $D_1$, and $n_S$ is another mono-sized packing of size $D_2 \neq D_1$. By Eq. 24, $\Phi_{\text{PSD}} = 0$, even when $D_2$ is nearly the same size as $D_1$, implying that the simulant provides "no value" for performing mechanical tests like drilling in the regolith, even though the two distributions provide almost identical mechanical behavior. Even when coarse binning with sieve screens, by Eq. 25 $\Phi_{\text{PSD}} = 0$ if $D_1$ and $D_2$ happen to fall on different sides of one of the sieve screens. Generalizing this problem, consider a particle size distribution composed of a set of discrete particle sizes. Murray et al. (2012) studied the transport coefficients for granular materials in kinetic theory and showed that the errors are vanishing when lunar soil is represented by a set of only *N*=8 discrete particle sizes in selected proportions, meaning that the regolith with a discretized particle size distribution can have mechanical behaviors indistinguishable from a regolith with a continuous size distribution. This is a useful result for computer modeling of regolith or for constructing a simulant by mixing a small number of monosized materials. However, $\Phi_{PSD}$ will calculate as approximately zero per Eq. 24 with such a simulant. The general problem is that the (pseudo) inner product calculated in Eq. 24 treats each particle size as an independent dimension in the $\ell_1$ space, and in linear algebra there is no concept of the "nearness" of one dimension to one another and thus no credit is given when $D_2 \approx D_1$. There is a deep literature on other mathematical methods besides vector space arithmetic to determine how similar two signals are to one another.

Method 2

We previously (Metzger et al., 2018) considered calculating $\Phi_{PSD}$ simply as a scalar comparison of power indices, similar to the scalar calculations for $\Phi_D$ and $\Phi_{BD}$, writing it as,

$$\Phi_{PSD} = \max\left\{0, 1 - \frac{1}{w}\frac{|q_S - q_R|}{|q_R|}\right\} \quad (26)$$

We measured $F^S_{\leq}(D_i)$ of the CI simulant using a CILAS Particle Size Analyzer model 1080. This identified $D_{min} = 0.01$ μm and $D_{max} = 170$ μm as the size range of the simulant. We converted the data points to the differential number of particles by discrete differentiation,

$$n_S(D_i) = \frac{F^S_{\leq}(D_i) - F^S_{\leq}(D_{i-1})}{D_i - D_{i-1}} \left(\frac{2}{D_i + D_{i-1}}\right)^3 \quad (27)$$

We found that $n_S$ for this CI simulant is approximated by a power law with $q_S = -3.44$ over three logarithmic decades, as shown in Fig. 7.

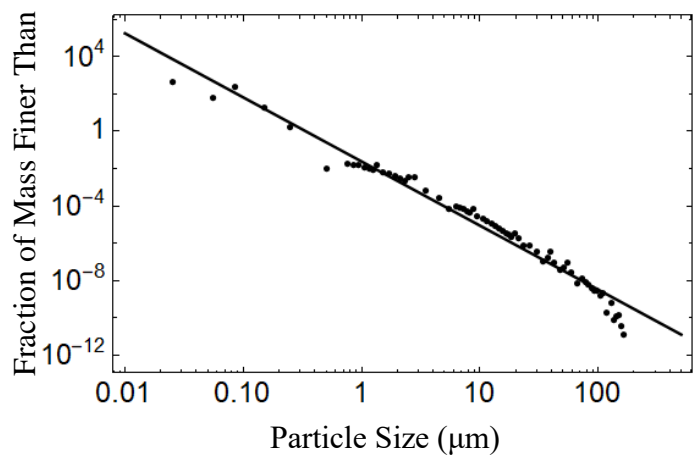

**Figure 7.** Differential particle number distribution for the CI Simulant (dots) with power law $q_S = -3.44$.

Alternatively, the equation for a power law $n(D) = c\,D^q$ with arbitrary $q$ may be converted to cumulative mass fraction form,

$$F_{\le}(D) = \frac{\int_{D_{min}}^{D} n(D')D'^3 dD'}{\int_{D_{min}}^{D_{max}} n(D')D'^3 dD'} \tag{28}$$

$$F_{\le}(D) = \begin{cases} \max\left(0,\ \min\left(1,\ \frac{D^{4+q}-D_{min}^{4+q}}{D_{max}^{4+q}-D_{min}^{4+q}}\right)\right), & \text{if } q \ne 3 \\ \max(0,\ \min(1,\ D/D_{max})), & \text{if } q = 3 \end{cases} \tag{29}$$

and this function may be least-squares fitted onto the simulant data $F^S_{\le}(D_i)$, finding $D_{min} = 1.0$ μm, $D_{max} = 105.0$ μm, and $q_S = -3.64$ as shown in Fig. 8. These two fitting methods returned different values of $q_S$ ($-3.44$ vs. $-3.64$) because the simulant is not perfectly a power law, although it is a good approximation.

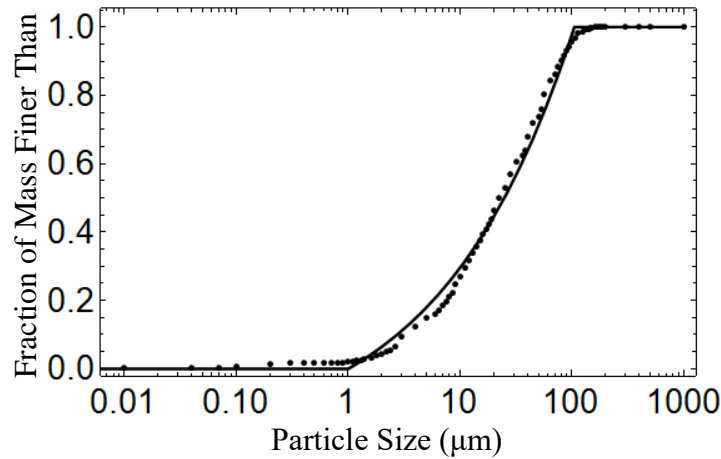

**Figure 8.** Dots: cumulative mass fraction of CI simulant. Solid line: least squares fit power law (converted to cumulative mass fraction), with best-fit parameters $D_{min} = 1.0$ μm, $D_{max} = 105.0$ μm, and $q_S = -3.64$.

Using Eq. 29 with $w = 1$ and $q_R = -3.5$, $q_S = -3.44$ yields $\Phi_{PSD}^{(old)} = 0.98$ or $q_S = -3.64$ yields $\Phi_{PSD}^{(old)} = 0.96$. While this is simple to calculate, it fails to quantify how far the simulant deviates from its own best fit power law. The UCF/DSI-CI-2 simulant does closely follow a power law, but other simulants might not. Therefore, this method does not make a generally valid FoM.

Method 3

A better method to calculate $\Phi_{PSD}$ is similar to the one adopted for $\Phi_{VR}$, to consider cumulative distributions $F^S_\leq(D)$ and $F^R_\leq(D)$ and perform an integral over the magnitude of their difference,

$$\delta F = \int_0^\infty \left| F^S_\leq(D) - F^R_\leq(D) \right| d(\log_{10} D) \tag{30}$$

$$\Phi_{PSD} = \max\left\{0, 1 - \frac{\delta F}{\log_{10} w}\right\} \tag{31}$$

where $F^S_\leq(D)$ is an analytical fit or piecemeal linear interpolation between the measured data points of the simulant, and $w$ is a scaling parameter that will be chosen to give reasonable results. In measuring $F^S_\leq(D)$ it is important that the data points be spaced sufficiently close together (e.g., closely spaced sieve screen sizes) that any features of the curve such as dips or spikes do not escape detection. By integrating the area between the two cumulative distributions, this formulation penalizes the simulant for having too many or two few particles at a given particle size, but the penalty is mitigated if the two curves quickly come back together again because the simulant provided the particles at nearly the correct particle size. Unlike the formulation of $\Phi_{VR}$, here we chose logarithmic integration to give equal weight to each logarithmic decade of the particle size range rather than giving equal weight to each micron of size. The fines contribute most of the surface area (cohesion) in regolith but the coarse particles contribute most of the mass (inertia and weight), so both ends of the spectrum should be significant in quantifying the fidelity of mechanical behavior. Eq. 30 can equivalently be written,

$$\delta F = \int_{-\infty}^\infty \left| F^S_\leq(10^n) - F^R_\leq(10^n) \right| dn \tag{31}$$

The reference model needs specificity in $D_{min}$ and $D_{max}$ to perform the integral, but they are inadequately constrained for asteroids and user-adjustable in the simulant. Therefore, the procedure for now is to supply the reference model with values of $D_{min}$ and $D_{max}$ corresponding to the particle size range of the simulant and to report them along with the FoM. There is also a pragmatic reason it must be done this way. For example, the term "soil" for the Moon is generally understood to include only particles up to about 2 mm diameter, although particles of all sizes including gravel, cobbles and larger are thoroughly mixed into the soil. Lunar simulants generally attempt to replicate only the "soil" size fraction up to 1 mm or 2 mm. Simulant users can add simulated gravel and cobbles into the simulant if needed for testing. NASA's original FoM method per Eqs. 17 and 18 was to calculate only up to this arbitrary $D_{max} = 1$ or 2 mm representing the size range of the simulant rather than the actual regolith, since it excludes the larger particles. In the future when asteroid particle sizes are better constrained, $D_{min}$ and $D_{max}$ will be defined as part of the reference model but the FoM will still need to be calculated over a more limited size range relative to the simulants.

We adopt this third method and illustrate it using the CI simulant. As shown in Figs. 7 and 8, the simulant has a power law distribution, $q_S = -3.44$ or $-3.64$, over a significant particle size range closely matching the power index of the bulk regolith reference model, $q_R = -3.5$. Because of this we expect the $\Phi_{PSD}$ for bulk regolith to be very high but the $\Phi_{PSD}$ for surficial regolith $q_R = -2.5$ to be much lower. We calculated both values of $\Phi_{PSD}$ per Eqs. 24 and 24A using the measured $F^S_\leq(D_i)$ with linear interpolation between data points, using $D_{min} = 0.01$ μm and $D_{max} = 170.0$ µm for size range of the reference model in Eq. 29 as this is the full size range of the simulant. We used $w = 3.5$, which will be justified below. This resulted in $\Phi_{PSD} = 0.55$ for $q_R = -3.5$ and $\Phi_{PSD} = 0.0$ for $q_R = -2.5$. The former is significantly less than 1 because the simulant departs from the power law for $D \lesssim 0.01$ µm and $D_{max} \gtrsim 105.0$ µm. If we define the reference model over just this narrower size range, the FoM values increases, indicating that the simulant is more useful for mechanical tests if the narrower range of particle sizes is important. For example, a spacecraft may pass regolith through a sieve into an inlet chute to an oven, so only the size range passing the sieve matters in the simulant tests. Taking this further, $\Phi_{PSD}$ can maximized for $q_R = -3.5$ when the reference model is restricted to $D_{min} = 0.81$ µm and $D_{max} = 86.2$ µm producing $\Phi_{PSD} = 0.89$, or for $q_R = -2.5$ when $D_{min} = 0.01$ µm and $D_{max} = 42.5$ µm yielding $\Phi_{PSD} = 0.57$. Fig. 9 shows $F^R_\leq(D)$ for both bulk and surficial regolith defined over these restricted size ranges compared to $F^S_\leq(D)$. Since $\Phi_{PSD}$ is necessarily a function of the reference model's size range (not just a function of $q_R$), and since size range requirements differ with the intended application, we recommend that particle sizing FoMs be reported as exemplified in Table 8. Providing less information than this would probably be inadequate.

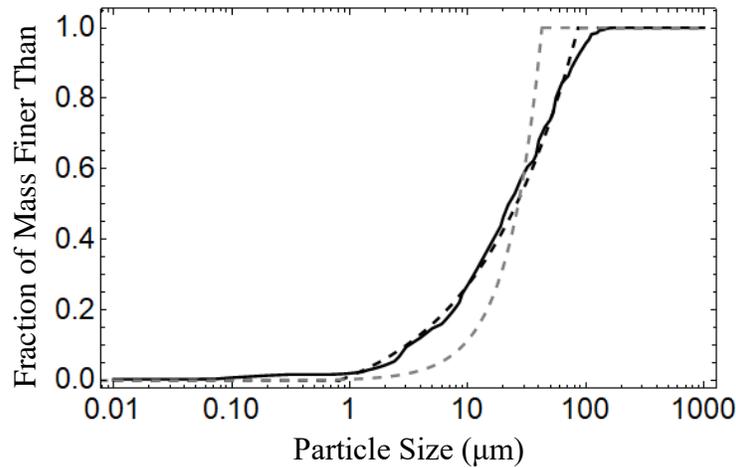

**Figure 9.** Solid Black: cumulative mass fraction $F^S_\leq(D)$ of the CI Simulant using linear interpolation between measured data points. Dashed black line: cumulative mass fraction reference model $F^R_\leq(D)$ per Eq. 29 with $q_R = -3.5$, $D_{min} = 0.81$ µm and $D_{max} = 86.2$ µm, which yields $\Phi_{PSD} = 0.89$. Dashed gray line: $F^R_\leq(D)$ per Eq. 29 with $q_R = -2.5$, $D_{min} = 0.01$ µm and $D_{max} = 42.5$ µm, which yields $\Phi_{PSD} = 0.57$.

**Table 8.** Example of Recommended Information to Report for $\Phi_{PSD}$

| Reference Material or Reference Model | Over Particle Size Range (μm) | Particle Size FoM |
|---|---|---|
| Bulk Regolith Reference Model, $q_R = -3.5$ | 0.01 – 170 (full range) | 0.55 |
| | 0.81 – 86.2 (best fit) | 0.89 |
| Surface Regolith Reference Model, $q_R = -2.5$ | 0.01 – 170 (full range) | 0.00 |
| | 0.01 – 42.5 (best fit) | 0.57 |

The goal in selecting the value of $w$ was to scale $\Phi_{PSD}$ so it reasonably represents the mechanical usefulness of an asteroid simulant in engineering tests. $\Phi_{PSD} = 0$ should communicate that the simulant is so dissimilar to the target regolith that mechanical tests using the simulant probably have "no engineering value" in predicting what will happen when the spacecraft reaches the asteroid, while $\Phi_{PSD} = 1$ means the simulant test should so adequately represent the asteroid (all other relevant FoMs being sufficiently high) that the engineering test is as valuable as we could hope, with linear scaling between. At present, we have only experienced judgement to rely upon in selecting $w$. In the future, it would be beneficial if experiments are performed to empirically test this selection. Fig. 10 shows values of $\Phi_{PSD}$ for the CI simulant as both $q_R$ and $w$ are varied. If $w$ is chosen to be too small, then $\Phi_{PSD}$ predicts too strictly that a test has no value even though the power index $q_R$ is only a little different than $q_S$. On the other hand, if $w$ is chosen to be too large, then $\Phi_{PSD}$ can say a test is valuable even though $q_R > -1$ (regolith vastly more coarse than the simulant) or $q_R < -5$ (regolith vastly more fine than the simulant). Fig. 11 shows the simulant data with seven hypothetical reference models (each representing the regolith on a hypothetical asteroid) per Eq. 29 with values of $q_R$ between -0.79 and -4.21. These reference models each use $D_{min} = 0.81$ μm and $D_{max} = 86.2$ μm which are the values that produced maximum $\Phi_{PSD}$ for $q_R = -3.5$. The values of $\Phi_{PSD}$ annotated on the plot are for the CI Simulant relative to each of these reference models using $w = 3.5$. The two extremal cases are for values of $q_R$ right at the limit of "no value", $\Phi_{PSD} = 0$. The finer of these, $q_R = -4.21$, has about 67% of its mass in the form of fine dust $D < 10$ μm, compared to the simulant having only 27% of its mass that fine. The coarser of these, $q_R = -0.79$, has over 80% of its mass coarser than 50 μm, while the simulant has only about 26% that coarse. It is a reasonable judgement that a drilling test in the CI simulant would provide essentially no useful information on how the drill would perform in either of those materials beyond what we could guess without doing the test. Therefore, it is reasonable for the FoM to become zero at approximately those boundaries as set by $w = 3.5$. The 50% value and 75% value curves are also shown in the plot, and they reasonably agree with intuition about mechanical performance and similarity of the particle size distributions. This demonstrates w = 3.5 is an acceptable scaling value.

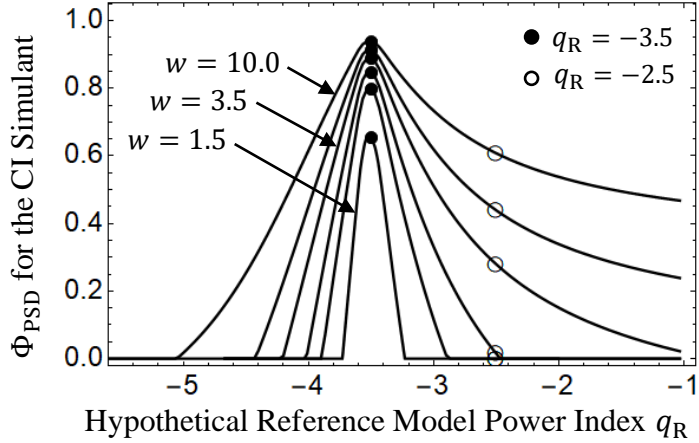

**Figure 10.** $\Phi_{PSD}$ of the CI simulant calculated for different (hypothetical) reference model power indices $q_R$ to evaluate the scale factors $w = 10.0$, 5.0, 3.5, 2.5, 2.0, and 1.5 (top to bottom). Filled and open circles represent bulk and surficial asteroid regolith, respectively.

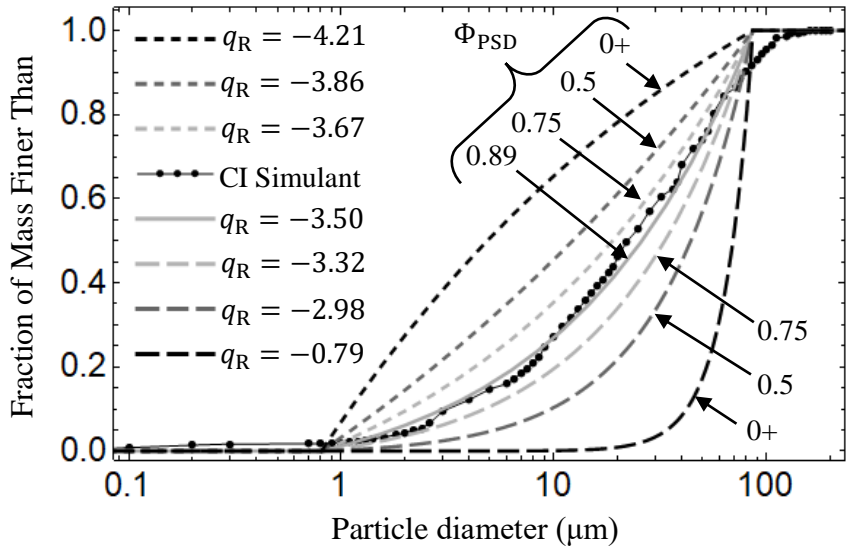

**Figure 11.** $F^R_{\leq}(D)$ for several hypothetical reference models of varied $q_R$, sharing the same $D_{max}$ and $D_{min}$, each representing a hypothetical asteroid. The annotated $\Phi_{PSD}$ values were calculated for each case versus the simulant $F^S_{\leq}(D)$ (dots) using $w = 3.5$. For predicting performance of a mechanical process like drilling on that asteroid, "$\Phi_{PSD} = 0.75$" means that a mechanical test with the CI Simulant has 75% of the value of a test with a material that exactly matches that asteroid per the reference model. The $\Phi_{PSD}$ scores seem appropriate for each curve, indicating $w = 3.5$ is a reasonable choice. Curves labeled $\Phi_{PSD} = 0+$ indicate the limits outside of which $\Phi_{PSD} = 0$.

## XI. Discussion

The eight calculated FoMs for the CI simulant are summarized in Table 9. To put these results into perspective, Table 10 compares the mineralogical FoMs for the CI simulant and several lunar simulants measured by Schrader et al. (2008). Some lunar simulants were created only to be physical simulants, matching particle sizing and grain density but not mineralogy or magnetic properties. For simulating bulk asteroid regolith UCF/DSI-CI-2 is an all-purpose simulant suitable for a wide variety of tests including drilling, resource extraction, magnetic anchoring, and radiation shielding. The volatile release FoM is a lower, because the simulant releases a lower mass than the reference meteorite. Users of the simulant will need to take that into account when performing volatile extraction tests.

Some of the new asteroid FoMs have little or no application to lunar simulants. $\Phi_{VR}$ may be of limited value for lunar simulants since most lunar soils contain very little hydrated minerals and make no attempt to simulate solar wind implanted volatiles. $\Phi_{BD}$ may be impractical for the Moon since it is a differentiated body and there are cobbles of many different mineralogies and thus a single FoM for bulk density is insufficient. On the other hand, the newly developed $\Phi_E$ and magnetic $\Phi_{MS}$ would be beneficial if applied to lunar simulants, expanding the set of lunar FoMs beyond the current four.

Our future work includes completing development of the five simulants listed in Table 1 and calculating the set of eight FoMs for each. These are all carbonaceous asteroid types, but they represent different petrological types and have a wide range of mechanical strengths, so this should provide additional tests of the usefulness of this system of FoMs. Insofar as possible, we intend to apply these FoMs to the available Martian and lunar simulants and other extant simulants. We have published an online database of simulants (http://sciences.ucf.edu/class/planetary-simulant-database) where we intend to provide a Fit-to-Use Table for all the above.

NASA's objectives in developing the FoM system were (1) to improve communication about simulants to improve effectiveness of the research community, (2) to provide higher confidence in test results when validating mission hardware for flight, and (3) to help simulant users avoid the common misuses of simulants that occurred in the lunar community. To help accomplish these objectives, we ask the community to participate in evolving the reference models, to provide any critique and improvements necessary of the FoM formulations, and to begin using them to characterize and report the simulants for both scientific and engineering research.

**Table 9.** Figures of Merit for CI Asteroid Simulant

| Figure of Merit | Reference Sample or Model | Calculated FoM |
|---|---|---|
| Mineralogical | Orgueil analyzed by Bland et al. (2004) | 0.83* |
| Elemental | Stoichiometry from the mineral analysis of Orgueil | 0.94* |
| Mineral Grain Density | Measurements of Orgueil by Consolmagno and Britt (1998) | 0.75 |
| Cobble Bulk Density | Measurements of Orgueil by Consolmagno and Britt (1998) | 0.72 |
| Magnetic Susceptibility | Multiple measurements of Orgueil | 0.96 |
| Cobble Strength | Model based on micro-indenter measurements of tensile strength of Orgueil by Tsuchiyama et al. (2008) scaled-up to compressive strength of larger fragments | 0.77 |
| Volatile Release | Orgueil analyzed by King et al. (2015) | 0.53 |
| Particle Sizing | Model based on Hayabusa sample return, boulder counting on Itokawa, and disrupted asteroids | 0.55 – 0.89** (bulk) 0 – 0.57** (surficial) |

*Preliminary based on vendor source data
** Depending on range of particle sizes; see text.

**Table 10.** Comparing Mineralogical Figures of Merit for Lunar and Asteroid Simulants

| Simulant | Reference Material | Figure of Merit |
|---|---|---|
| UCF/DSI-CI-2 | Orgueil | 0.83* |
| NU-LHT-1M | Lunar 64001/64002 | 0.65 |
| NU-LHT-2M | Lunar 64001/64002 | 0.55 |
| OB-1 | Lunar 64001/64002 | 0.28 |
| JSC-1 | Lunar 64001/64002 | 0.33 |
| JSC-1A | Lunar 64001/64002 | 0.35 |
| JSC-1AF | Lunar 64001/64002 | 0.43 |
| MLS-1 | Lunar 64001/64002 | 0.35 |
| FJS-1 | Lunar 64001/64002 | 0.36 |

*Preliminary based on vendor source data

### XII. Summary and Conclusions

We have applied and extended NASA's Figure of Merit system for lunar regolith simulants by developing eight Figures of Merit for asteroid simulants, which are summarized in Table 11. Each FoM is a mathematical formula to compare one property of an asteroid simulant (regolith and/or cobble variety) to the same property in a reference meteorite or asteroid. For the regolith particle size distribution FoM, we were unable to use meteorites for the reference data so we instead developed a reference model. For the cobble mechanical strength FoM, the available data on strengths of certain types of meteorites are inadequate so we built a reference model upon the available data. We illustrated use of these FoMs and reference models by calculating their values for the UCF/DSI-CI-2 simulant that was recently developed for NASA. The set of FoMs adequately quantified the fidelity of the simulant's behaviors for many tests needed by the user community. Future work will measure FoMs for the other asteroid spectral class simulants currently under development and, insofar as possible, for lunar, Martian, and other available simulants. Their values will be summarized in a Fit-to-Use Table that tells which simulants are best for which types of tests representing different extraterrestrial terrains. We recommend that the research and technology development communities begin using and improving the FoM system to gain the benefits NASA intended when initiating this system.

**Table 11.** Summary of Asteroid Simulant Figures of Merit

| FoM | Property | Type | Equation | $w$ |
|---|---|---|---|---|
| $\Phi_M$ | Mineralogical | Vector (pseudo-) inner product in $\ell_1$ | $\Phi_M(\vec{S}_M, \vec{R}_M) = \|\vec{S}_M \cap \vec{R}_M\|_1 = \sum_{i=1}^{N_M} \min(s_i, r_i)$ | -- |
| $\Phi_E$ | Elemental | Vector (pseudo-) inner product in $\ell_1$ | $\Phi_E(\vec{R}_E, \vec{S}_E) = \|\vec{S}_E \cap \vec{R}_E\|_1 = \sum_{i=1}^{N_E} \min(v_i, w_i)$ | -- |
| $\Phi_D$ | Average Grain Density (cobbles and regolith) | Linear scalar difference | $\Phi_D = \max\left\{0, 1 - \frac{1}{w}\frac{\left|\rho_g^S - \rho_g^R\right|}{\rho_g^R}\right\}$ | 0.5 |
| $\Phi_{BD}$ | Cobble Bulk Density | Linear scalar difference | $\Phi_{BD} = \max\left\{0, 1 - \frac{1}{w}\frac{\left|\rho_{bulk}^S - \rho_{bulk}^R\right|}{\rho_{bulk}^R}\right\}$ | 0.5 |
| $\Phi_{CS}$ | Cobble strength | Logarithmic scalar difference | $\Phi_{CS} = \max\left\{0, 1 - \frac{\left|\log_{10} U^S - \log_{10} U^R\right|}{\log_{10} w}\right\}$ | 5 |
| $\Phi_{MS}$ | Magnetic Susceptibility | Logarithmic scalar difference | $\Phi_{MS} = \max\left\{0, 1 - \frac{\left|\log_{10} \chi^S - \log_{10} \chi^R\right|}{\log_{10} w}\right\}$ | 5 |
| $\Phi_{VR}$ | Volatile Release | Linear $\ell_1$-norm of difference function | $\Phi_{VR} = \max\left\{0, 1 - \frac{1}{w}\frac{\int_{T_{min}}^{T_{max}}|v_S - v_R|\,dT}{\int_{T_{min}}^{T_{max}}|v_R|\,dT}\right\}$ | 1 |
| $\Phi_{PSD}$ | Particle Size Distribution | Logarithmic $\ell_1$-norm of difference function | $\Phi_{PSD} = \max\left\{0, 1 - \frac{\int_{\log D_0}^{\infty}\left|F^S_{\leq}(D) - F^R_{\leq}(D)\right|\,d(\log D)}{\log_{10} w}\right\}$ | 3.5 |

**Acknowledgements**

This work was directly supported in part by NASA's Solar System Exploration Research Virtual Institute cooperative agreement award NNA14AB05A, and it was supported in part by contract number NNX15CK10P, "Task-Specific Asteroid Simulants for Ground Testing," as part of NASA's Small Business Innovative Research (SBIR) program. Declaration of Interest: Stephen Covey is employed by Deep Space Industries, commercial provider of asteroid simulant UCF/DSI-CI-2.

45. Lee, P. (1996). "Dust levitation on asteroids." *Icarus*, 124(1), pp.181-194.

46. Liu, Yang, Lawrence A. Taylor, James R. Thompson, Darren W. Schnare, and Jae-Sung Park. "Unique properties of lunar impact glass: Nanophase metallic Fe synthesis." *American Mineralogist* 92, no. 8-9 (2007): 1420-1427.

47. Lunar Exploration Analysis Group (LEAG). "Status of Lunar Regolith Simulants and Demand for Apollo Lunar Samples." Report to the NASA Advisory Council. http://www.lpi.usra.edu/leag/reports/SIM_SATReport2010.pdf. 2010.

48. Macke, Robert J., Guy J. Consolmagno, and Daniel T. Britt. "Density, porosity, and magnetic susceptibility of carbonaceous chondrites." *Meteoritics & Planetary Science* 46, no. 12 (2011): 1842-1862.

49. Makabe, Teruo, and Hajime Yano. "The effective projectile shape for asteroid impact sampling." In *Proceedings of the 26th International Conference on Space Technology and Science*, paper no. 2008-k-08. 2008.

50. Mazrouei, S., M. G. Daly, Olivier S. Barnouin, C. M. Ernst, and I. DeSouza. "Block distributions on Itokawa." *Icarus* 229 (2014): 181-189.

51. McKay, David S. and James D. Blacic (1991). *Workshop on Production and Uses of Simulated Lunar Materials*. LPI Tech. Rpt. 91-01. Lunar and Planetary Institute, Houston, TX. 83 pp.

52. McKay, D. S., B. L. Cooper, and L. M. Riofrio. "New measurements of the particle size distribution of Apollo 11 lunar soil." In *Lunar and Planetary Science Conference*, vol. 40. 2009.

53. McKay, D. S., R. M. Fruland, and G. H. Heiken. "Grain size and the evolution of lunar soils." In *Lunar and Planetary Science Conference Proceedings*, vol. 5, pp. 887-906. 1974.

54. McLemore, Carole A. "Logic of the NASA/MSFC Simulant Development Technical Approach." http://isru.msfc.nasa.gov/simulantdev_logic.html (Updated 20 June 2014).

55. McSween Jr, Harry Y., David W. Mittlefehldt, Andrew W. Beck, Rhiannon G. Mayne, and Timothy J. McCoy. "HED meteorites and their relationship to the geology of Vesta and the Dawn mission." In *The Dawn Mission to Minor Planets 4 Vesta and 1 Ceres*, pp. 141-174. Springer New York, 2010.

56. Metzger, Philip T., and Daniel T. Britt, "Model for Asteroid Regolith to Guide Simulant Development," submitted to *Icarus* (2018).
37

79. Schrader, Christian, Doug Rickman, Carole Mclemore, John Fikes, Stephen Wilson, Doug Stoeser, Alan Butcher, and Pieter Botha. "Extant and extinct lunar regolith simulants: Modal analyses of NU-LHT-1M and-2m, OB-1, JSC-1, JSC-1A and-1AF, FJS-1, and MLS-1." (2008).

80. Schrader, Christian M., and Douglas L. Rickman. "Overview of figure of merit analyses of simulants and the fit-to-use matrix." MSFC-2222, Lunar Regolith/Simulant Workshop; 17-20 Mar. 2009; Huntsville, AL (2009).

81. Schrader, Christian M., Douglas L. Rickman, Carole A. Mclemore, John C. Fikes, Douglas B. Stoeser, Susan J. Wentworth, and David S. McKay. "Lunar regolith characterization for simulant design and evaluation using figure of merit algorithms." In *Proc., 47th AIAA Aerospace Sciences Meeting. ISO*, vol. 690. 2009.

82. Schrader, C. M., D. L. Rickman, C. A. McLemore, and J. C. Fikes. "Lunar Regolith Simulant User's Guide." NASA Technical Memorandum 2010–216446, NASA Marshall Space Flight Center, Ala. (2010).

83. Sibille, Laurent, Paul Carpenter, Ronald Schlagheck, and Raymond A. French. "Lunar regolith simulant materials: recommendations for standardization, production, and usage." NASA Technical Publication 2006–214605, NASA Marshall Space Flight Center, Ala (2006).

84. Sickafoose, A.A., Colwell, J.E., Horányi, M. and Robertson, S. (2002). Experimental levitation of dust grains in a plasma sheath. *J Geophys Res: Space Phys*, *107*(A11).

85. Snodgrass, C., Tubiana, C., Vincent, J.B., Sierks, H., Hviid, S., Moissi, R., Boehnhardt, H., Barbieri, C., Koschny, D., Lamy, P. and Rickman, H., 2010. A collision in 2009 as the origin of the debris trail of asteroid P/2010A2. *Nature*, *467*, pp.814-816.

86. Stevenson, R., E. A. Kramer, J. M. Bauer, J. R. Masiero, and A. K. Mainzer, "Characterization of active main belt object P/2012 F5 (Gibbs): A possible impacted asteroid." *The Astrophysical Journal* 759 no. 2 (2012): 142.

87. Svetsov V. V., Nemtchinov I. V., and Teterev A. V. 1995. Disintegration of large meteoroids in Earth's atmosphere: Theoretical models. *Icarus* 116:131–153.

88. Taylor, Lawrence A., and Yang Liu. "Important considerations for lunar soil simulants." *Earth and Space 2010: Engineering, Science, Construction, and Operations in Challenging Environments* (2010).

89. Taylor L.A., Pieters C.M., and Britt D.T. (2016) Evaluations of lunar regolith simulants *Planetary and Space Science*, Volume 126, July 2016, Pages 1–7.
40